# Unbundling Transaction Services in the Cloud

David Lomet
Microsoft Research
Redmond, WA 98052
+01-425-703-1853
lomet@microsoft.com

Alan Fekete
University of Sydney
Sydney, Australia
+61-2-93514287
a.fekete@usyd.edu.au

Gerhard Weikum
Max Planck Institute
Saarbrücken, Germany
+49-681-9325-500
weikum@mpi-sb.mpg.de

Mike Zwilling
Microsoft Corporation
Redmond, WA 98052
+01-425-703-6378
mikezw@microsoft.com

## ABSTRACT

The traditional architecture for a DBMS engine has the recovery, concurrency control and access method code tightly bound together in a storage engine for records. We propose a different approach, where the storage engine is factored into two layers (each of which might have multiple heterogeneous instances). A Transactional Component (TC) works at a logical level only: it knows about transactions and their "logical" concurrency control and undo/redo recovery, but it does not know about page layout, B-trees etc. A Data Component (DC) knows about the physical storage structure. It supports a record oriented interface that provides atomic operations, but it does not know about transactions. Providing atomic record operations may itself involve DC-local concurrency control and recovery, which can be implemented using system transactions. The interaction of the mechanisms in TC and DC leads to multi-level redo (unlike the repeat history paradigm for redo in integrated engines). This refactoring of the system architecture could allow easier deployment of application-specific physical structures and may also be helpful to exploit multi-core hardware. Particularly promising is its potential to enable flexible transactions in cloud database deployments. We describe the necessary principles for unbundled recovery, and discuss implementation issues.

## Categories and Subject Descriptors
**H.2.4 [Systems]:** Concurrency, Transaction processing

**H.2.2 [Physical Design]:** Recovery and restart, access methods

## 1. INTRODUCTION

DBMS decomposition has been suggested by several researchers [2, 8, 21], but has remained an elusive goal, "up in the clouds", for two decades. One can indeed easily separate the query processing and optimization components from the storage engine. However, as observed in [10], "The truly monolithic piece of a DBMS is the transactional storage manager that typically encompasses four deeply intertwined components:

1. A lock manager for concurrency control.
2. A log manager for recovery.
3. A buffer pool for staging database I/Os.
4. Access methods for organizing data on disk."

Folk wisdom, beginning with System R [6, 7], suggests that this integration is a requirement for high performance from these system elements, since they are exercised continuously during DBMS execution. Nevertheless, cloud computing re-introduces interest in and pressure for again tackling this challenge of unbundling transaction services and data management.

### 1.1 Industry Trends
Trends within the computing systems industry, especially for database systems, require us to rethink the database systems architecture and to consider disentangling the previously integrated aspects of the database kernel, transactional services going to a transactional component (TC) that is architecturally separate from data services (access methods and cache management) in a data component (DC). These imperatives are:

1. Cloud computing opens up opportunities for easy deployment of new, perhaps application dependent, database management. Cloud deployments create new problems of scale and computing infrastructure. Separating TC functionality from DC functionality enables cloud platforms to support transactions with much greater flexibility, regardless of where in the cloud the data and its DCs reside.

2. New, light-weight data-management engines for specific application areas ("one size does not fit all" [22]) call for a composable run-time infrastructure with low overhead. For example, one might build an RDF engine as a DC with transactional functionality added as a separate layer.

3. The major hardware trends of our time are (1) increasing numbers of cores on processor chips, and (2) increasing main memory latency. This suggests a rethinking of database architecture (even for traditional database applications such as OLTP [9]) to enhance parallelism and improve cache hit ratios. The decomposition into TC and DC may improve both processor (core) utilization since each component could run on a separate core, and cache performance, since each component will have shorter code paths and may result in much higher hit rates for the instruction cache(s) of one core.

4. Substantial processing power has existed for many years within the controllers for I/O subsystems. One appealing notion has been to move part of database functionality out of the cpu and into these controllers. Separating the data component as we are suggesting, permits moving this "data centric" functionality to the storage controllers, enabling a "disk" to support a record oriented interface instead of a page oriented interface.

5. A classic goal has been extensible database management systems. Adding a new access method to support new data types (e.g., shapes, avatars, etc. used in virtual worlds, for games, and 3D Internet) and their associated search needs is eased substantially when the type implementation (as DC) can rely on transactional services provided separately by TC.

## 1.2 Our Contribution

What makes partitioning a database kernel difficult is that state-of-the-art concurrency control and recovery relies on knowledge of the way that storage is paginated, and how records are assigned to pages. For example, physiological logging [6, 24] requires each log record to deal with a single page. Also, state-of-the-art access methods use sophisticated ways to get high concurrency.

Our contribution is an architecture for database kernels in which transactional functionality in a TC is unbundled from the access methods and cache management in a DC. The TC does all locking for transactional concurrency control and logging for transaction abort and durability. All knowledge of pages is confined to a DC, which means that the TC must operate at the logical level on records. The TC invokes (and logs) logical operations of a DC. This is pictured in Figure 1. Our design differs from [21], where access methods are done on top of a transactional layer.

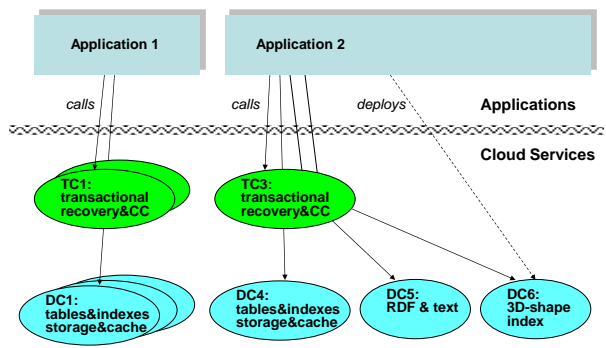

**Figure 1: Architecture of an unbundled database kernel.**

A DC knows nothing about transactions, their commit or abort. It is required to make the individual logical operations atomic and idempotent. Idempotence of DC operations permits the TC to resend operations to the DC, either during normal execution (perhaps after a response is lost) or later during recovery, while ensuring exactly-once execution of the overall system.

Both the TC and DC are multi-threaded, which is essential for high performance, but which introduces a number of subtle issues with which both TC and DC must deal. For example, TC has an obligation to never send logically conflicting operations concurrently to a DC. In this way, the order of logical log records written by the TC can be guaranteed to be consistent with the physical ordering performed in a DC.

Because a DC completely handles the pagination required for an access method like B-trees, it is the DC that must deal with page splits and deletes. Such structure modifications themselves require concurrency control and recovery. Integrating recovery across both transaction and access method levels is a characteristic of modern database systems [10], but providing them separately from each other requires thinking anew about multi-level recovery. We address this in Section 4.2.

Providing separate TC and DC permits us to instantiate these components in a number of new and interesting ways. Because DCs have longer execution paths, one might deploy a larger number of DC instances on a multi-core platform than TC instances for better load balancing. In a cloud environment, one would want DCs to be close to the data, while the TCs might have a much looser coupling. While multiple TCs must never send conflicting operations to a DC (because the order of operations will not be logged at the DC), it is nonetheless possible for TCs to share data, especially when DCs provide a versioning capability. Deploying TCs that can share DCs in this way enables our architecture to support some of the interesting cloud scenarios, without introducing the need for two phase commit.

## 2. APPLICATION PERSPECTIVE

In the Web 2.0 landscape, there are new applications that desire fast and easy deployment on a care-free platform. Such applications include social-community forums (e.g., sharing photos, videos, reviews, ratings, etc.), Internet-based long-running games with many players, and information mashups that compose value-added portals from blogs, news, and other Internet sources. Cloud services are intended to provide a convenient solution for such applications. Unbundling, as we suggest, can help Web 2.0 get fast transfer of original ideas into popular Internet sites.

As one example, consider a Web 2.0 photo-sharing platform. On first thought, this may seem simple, merely needing persistent storage for large files. But the application also must manage users and their accounts, photo ownerships and access rights for other users, thematic groups for photos and users, friendships and other interactions among users, and so on. This should be consistent under high update rates; so there is a significant OLTP aspect.

Photos are associated with annotations (tags) and reviews. This entails referential integrity constraints; corresponding operations must be guarded by transactions with appropriate scope. Reviews consist of natural-language text, and the application may have a non-standard index structure for this (e.g., for phrases that express opinions). Similarly, advanced visualizations of say the tag-cloud dynamics may require special data structures as well. Finally, imagine fancy functionality that finds photos of the same object (e.g., the Golden Gate Bridge) uploaded by different users and combines them into a 3D model which in turn would be made searchable using the latest index structures for geometric shapes.

Of course, all this rich data could be mapped onto relational tables provided by a DBMS-style cloud service. But then the application could not utilize its advanced indexes for text phrases, 3D models, etc. Alternatively, it could use a simpler storage service, offered in the cloud, without transaction management. This service would just provide persistent store, with unlimited scalability and de-facto perfect availability, and the application program would implement its index structures on top of it. But now the application would also have to implement its own transactional concurrency control and recovery. The authors of [3] have shown how to do this with overhead; but we can do better by unbundling the transactional issues from the actual data management. The photo-sharing application could use a combination of already available file and table managers and home-grown index managers as DCs. For transaction management it could directly use the services of a TC, offered in the cloud. This TC (or these TCs if we instantiate it multiple times for scaling up throughput) would be able to interact with the various DCs via interaction contracts as will be explained in later sections. There is no free lunch, though. The application's home-grown DCs would have to be written so as to satisfy the DC parts of the contracts. This is simpler than designing and coding a high-performance transactional storage subsystem.

## 3. TECHNICAL CHALLENGES

The existing industry-standard solutions to concurrency control and recovery do not work when transaction services are separated from access methods and cache management. Most of this paper is focused on dealing with recovery issues, which require a larger departure from current practice. However, we first outline how we would deal with concurrency control differences as well.

### 3.1 Concurrency Control

For many operations, splitting the database kernel into TC and DC causes little trouble for a two phase locking approach to transactional concurrency control. The operations that involve updating or reading of records that are named by record identifiers can easily lock these records within a TC, prior to the TC sending the request to the DC that accesses the page containing the record. It is harder when ranges of records are being locked.

In existing systems where the database kernel is one integrated piece, a requested operation is actually executing within the page containing the data. Thus an operation dealing with a range can determine the keys involved, then lock them using, e.g., key range locking [13, 20], before performing the actual access. But in our unbundled approach, the TC needs to do the locking prior[1] to sending a request to the DC. That is, the lock must be obtained before it is known which keys are present in (or just after) the range. Thus we need to invest our lock manager and the TC code that uses it with techniques for locking ranges. We know of two ways to deal with the locking of ranges of records.

**Fetch ahead protocol:** Do an initial speculative probe to have the DC return the keys to the next (in order) collection of keys. At this point, the TC can lock those records, and submit the next request to do the read or write, together with a speculative request for the following keys. Should the records to be read or written be different from the ones that were locked based on the earlier request, this subsequent request becomes again a speculative request for the earlier records.

**Range locks:** Introduce explicit range locks that partition the keys of any table. Many systems currently support some form of this by permitting table locks or page locks, however our separation precludes us from locking pages. Each range of the partition is locked prior to accessing the requested records. There can be speculative record accesses at partition boundaries, but most accesses can proceed without this. This protocol avoids key range locking, and hence gives up some concurrency. However it should also reduce locking overhead since fewer locks are needed.

---

[1] This is to enforce the requirement that the DC never have two conflicting operations executing concurrently.

Either of these approaches can be made to work, so we now turn our attention to dealing with recovery.

### 3.2 Recovery

In an unbundled kernel, ARIES-style recovery [19, 20] does not work, even augmented with the usual multi-level recovery [14, 23] with physical repeating-history redo of log records, and logical, multi-level undo (which allows fine-grained concurrency control).

1. The DC provides only record-oriented *logical operations* where the TC knows nothing about pages. Hence, the TC log records cannot contain page identifiers. Redo needs to be done at a logical level. Pages and dealing with them is the exclusive province of the DC. Logical redo requires that, e.g. DC data structures be well formed (search correct) at the time that redo is performed, not simply when undo is performed While System R performed logical redo, it required operation-consistent checkpointing, which means that operation execution must be interrupted until no operations are active, at which point a checkpoint can be taken. This compromises both availability and performance.

2. *LSNs* are the normal way of ensuring operation *idempotence*. This is very convenient in the conventional setting where the LSN is assigned after a page is latched for update. With the TC doing the transactional logging, the situation is more complicated. The separation of the TC from the DC together with the independent *multi-threading* of TC and DC means that the TC will assign an LSN before the order in which operations access a page is determined. This can lead to *out-of-order executions* in which a later operation for a page with a higher LSN reaches the page before an earlier operation with a lower LSN. While these operations cannot conflict (see Section 2.1), the out-of-order LSNs must be dealt with.

3. DCs may *autonomously perform internal "system transactions"* (e.g., page splits and other index tree reorganizations) that might not commute with TC-initiated logical operations. Despite this, the DC needs to restore its indexes to a well-formed state prior to the TC executing recovery. Thus, the DC structure modification operations will execute during recovery out of their original execution order. Further, the TC has no way of knowing about these DC-internal actions. All it can do is assign LSNs and ensure that its redo repeats history by delivering operations in the correct order to the DC. The burden is on the DC to manage LSNs on pages in such a way that this TC strategy will work. Conventional techniques fail for this.

4. DC and TC may *independently fail*, and a crash of one of them should not force amnesia for the other component, e.g. by requiring the DC to discard all cached pages.

## 4. UNBUNDLED RECOVERY PRINCIPLES

We begin by describing the architecture of our separate TC and DC in terms of components which interact through exchange of particular messages. We then give the requirements on the interactions, to ensure that recovery can execute correctly. This is similar to our earlier work on recovery guarantees in distributed computing [1].

## 4.1 Database Kernel as a Distributed System

We envision the TC and the DC as two "distributed" components that have an arms-length interaction. We describe what the components do that are independent of each other, though both are important in providing a working system.

### 4.1.1 Transactional component (TC)

The TC acts as client to the DC. It wraps all requests to the kernel from higher in the database system or application stack. The TC needs to do the following:

1. Transactional locking to ensure that transactions are properly isolated (serializable) and that there are no concurrent conflicting operation requests submitted to the DC. The locks cannot exploit knowledge of data pagination.

2. Transaction atomicity, that is, ensuring that for every completed user transaction that is provided to TC from higher up the application stack, either

   a. The user transaction commits, after TC has caused DC to perform all the individual logical operations necessary to achieve the intended effect of the transaction, or

   b. The user transaction aborts, after TC has caused DC to perform a collection of logical operations whose combined effect is rollback, so there is no net change to the logical state. That is, TC must ensure that DC performs a (possibly empty) set of logical operations, followed in reverse chronological order by logical operations that are inverses of the earlier ones.

3. Transactional logging, both undo and redo, after appropriate locking. Undo logging in the TC will enable rollback of a user transaction, by providing information TC can use to submit inverse logical operations to DC. Redo logging in TC allows TC to resubmit logical operations when it needs to, following a crash of DC. That there are no conflicting concurrent operation requests ensures that *logical log records can be written in OPSR* (order-preserving serializable) order, even for actual out-of-order executions in multi-threaded mode. This must hold for whatever concurrency control method the TC chooses to use including fine-grained locking as well as optimistic methods.

4. Log forcing at appropriate times for transaction durability.

### 4.1.2 Data component (DC)

The DC acts as a server for requests from the TC. It is responsible for organizing, searching, updating, caching and durability for the data in the database. It supports a non-transactional, record oriented interface. The way in which the records are mapped to disk pages is known only to the DC itself, and is not revealed to the TC. It needs to do the following:

1. Provide atomic operations on its data (relational records, XML documents, encapsulated objects, etc.). Atomicity for individual logical operations is a form of linearizing concurrent operations [11], conceptually isolating them so that they appear as if they were indivisible with regard to concurrent executions [12, 18, 24]. More precisely, operation atomicity means that there is a total order on all the operations, compatible with any externally observable order (where one operation has returned before another is requested for the first time) and compatible with the results returned (so each operation's result reflects the state produced by all the operations ordered before that operation). Atomic operations ensure that serial replay of operations during recovery is possible. To allow multi-threading within DC, while still having atomic operations, each operation will need to latch whatever pages it operates on, until the operation has been performed on all the pages. However, as with page latches in traditional storage engines, these latches are held for very short periods, and latch deadlocks are avoided via the ordering of latch requests.

2. Maintain indexes and storage structures behind the scenes. For simple storage structures, each record lies on a fixed page, and DC can maintain the indices easily. However, for a structure like a B-tree, where a logical operation may lead to re-arrangements that affect multiple physical pages, the maintenance of indices must be done using system transactions that are not related in any way to user-invoked transactions known to the TC; implementation of system transactions may involve their own concurrency control and recovery.

3. Provide cache management, staging the data pages to and from the disk as needed.

## 4.2 TC:DC Interactions

Our earlier work [1] described "interaction contracts" which ensure that both sender and receiver of a message would agree on whether the message was sent, independently of system or communication failures. The principles listed below have similar intent, but there are differences, especially as in an unbundled database kernel, many interactions are not made stable immediately, but rather caching is used extensively, with state made stable lazily.

**Causality:** Causality means, that the sender of a message remembers that it sent the message whenever the receiver remembers receiving the message. This must be true during normal execution (trivial to do with volatile execution state) as well as in the case that one or more parts of the system fail. It is causality that leads to the classical write-ahead logging protocol. Partial failures are possible, whereby either TC and/or DC fail. *To respond to partial failures in a high performance way requires new cache management techniques for the DC* (see Section 4.3 and also [17]).

**Unique request IDs:** The TC labels each of its operations with a unique, monotonically increasing request identifier (usually an LSN derived from the TC log). TC request IDs make it possible for the DC to provide idempotence.

**Idempotence:** The DC manages request IDs within its data structures so that it can decide when its state already reflects the execution of the request, and when it does not. It must ensure that it can successfully execute all unexecuted requests so as to achieve their original results, both during normal execution and during restart. *Providing idempotence in our setting is a substantial technological challenge requiring new techniques.* (See Section 4.1)

**Resend Requests:** The TC resends the requests until it receives some form of acknowledgment from the DC. TC resend with unique request ids, working with DC idempotence, enable **exactly-once execution** of logical operations.

**Recovery:** The TC makes all requests to the DC in terms of logical (record-oriented) operations. The DC index structures must be well-formed for redo recovery to succeed. The DC must recover its storage structures first so that they are well-formed, before TC can perform redo recovery, not simply before undo recovery. Thus, *system transactions need to be logged such that they can be executed during recovery out of their original execution order.* (See Section 4.2)

**Contract termination:** There needs to be a protocol between TC and DC that permits the guarantees for causality and idempotence to be released. For example, the TC will eventually refrain from resending operations during restart. This corresponds to checkpointing in a conventional kernel; it involves coordinating the stable part of the recovery log managed by the TC with the stable part of the database state managed by the DC. *This does not require new techniques but we must expose functionality at the TC:DC interface.*

### 4.2.1 The TC/DC API

Here we summarize the interface through which necessary information is passed between TC and DC. We present these as functions or methods of DC, to be invoked by TC; however we do not limit the implementation technology for information exchange, and indeed we expect that in a cloud environment asynchronous messages might be used with the request flowing in on direction, with a later reply in the reverse direction, while signals and shared variables might be more suited for a multi-core design. Also, while usually TC is driving each interaction, there are some situations where DC will need to spontaneously convey information to TC; for example, following a crash of DC, a prompt is needed so that TC will begin the restart function.

**perform_operation**. TC needs to provide DC with the information about the logical operation, including the operation name and arguments (among which is the table name and the key for the record involved, or description of a range of keys as discussed in Section 3.1), and also a unique identifier (which is typically the LSN from the TC-log record for this operation). Resends of the request can be characterized by re-use of the operation identifier. The eventual reply for this request includes the operation identifier so it can be correlated to the request, as well as the return value of the operation itself. Note that the information given to DC does not carry any information about the user transaction of which it is a part, nor does DC know whether this operation is done as forward activity, or as an inverse during rollback of the user transaction.

**end_of_stable_log**. An argument, *EOSL*, is the LSN for the last entry from the TC-log that has been made stable. DC knows that all operations with this operation identifier, or lower, will not be lost in a crash of DC, and so causality allows DC to make any such operation stable in DC. This function is how WAL is enforced in an unbundled engine. A traditional storage engine performs exactly the same check but without using messages to convey the information.

**checkpoint**. An argument, *newRSSP*, is an LSN to which TC wishes to advance its **redo scan start point**. DC will reply once it has made stable all pages that contain operations whose LSN is below *newRSSP*; this releases the contract requiring TC to be willing to resend these operations, and only at this point can TC actually advance its start point for replaying operations in subsequent restarts. DC may also proactively make pages stable, and could spontaneously inform TC that the RSSP can advance to be after a given LSN.

**low_water_mark**. This function informs DC that TC has received the response from every logical operation with LSN up to and including the argument *LWM*, and so DC can be sure that there are no gaps among the lower *LSN* operations which are reflected in cache pages. The use of this information is discussed in Section 5.1.2. Like end_of_stable_log, this is important for deciding when pages in DC's cache can be flushed to disk. Thus one might trade some flexibility in DC for simplicity of coding, by combining end_of_stable_log and low_water_mark into one function that simply informs DC of the operation id, for which it is safe to flush a page from the DC cache so long as the page contains no operation beyond this LSN.

**restart**. We describe this as a single complicated function, but in practice the information passed would probably be batched and conveyed in several messages. TC informs DC that restart is commencing, and that it must discard any information about operations with LSNs higher than the last one in the stable TC log (these operations would be lost forever; causality ensures that any such information is not yet stable in DC) ; also the restart function includes resending all operations on the stable TC-log from the redo scan start point onwards; after they have been applied by the DC (which itself happens after DC resets its state, see Section 5.3.2), then TC will send logical operations which are inverses for those operations of user transactions that need rollback; finally, once all have been applied in DC, DC can acknowledge completion of the restart function, allowing normal processing to resume. If DC fails, we assume an out-of-band prompt is passed to TC, so TC knows to begin restart.

## 5. NEW TECHNIQUES

In this section, we describe some novel techniques to deal with the new complexities of providing "unbundled" recovery.

## 5.1 Out-of-Order Operation Execution

### 5.1.1 Current Technique

Because of the arms length separation of TC from DC, and their multi-threading, TC operation requests can arrive at the code accessing data on a page in an order that differs from the order of TC request ids (LSNs). This undermines the usual recovery test for idempotence in which a log operation's LSN is compared to an LSN stored in the data page impacted by the operation. This traditional test is simply: *Operation LSN <= Page LSN*

When this test is true in a monolithic system where logical log records are produced (and given LSNs) during a critical section in which the page is modified, it means that the page contains the effects of the operation, and redo is prohibited for the logged operation. Otherwise, the operation must be re-executed and the page (along with its LSN) is updated.

Because of out-of-order execution in an unbundled system, this test is no longer suitable. If an operation $O_j$ with $LSN_j$ executes before an operation $O_i$ with $LSN_i$, and $LSN_i < LSN_j$, and the page is immediately made stable after $O_j$'s execution, it will then contain a page L$S$N equal to $LSN_j$. The traditional test will incorrectly indicate that $O_i$ results are included in the page, and that there is no need to re-execute $O_i$.

This difficulty could be solved by introducing record level LSNs, since updates are conflicting record operations, and conflicting operations cannot execute concurrently. However, this is very expensive in the space required. Hence we prefer a page LSN oriented solution.

### 5.1.2 Our New Technique

To deal with out-of-order execution, we introduce the notion of an abstract page LSN denoted as *abLSN*. We then generalize the meaning of <= so that our test, showing when redo is not required, become *Operation LSN <= Page abLSN*.

We describe how this is done here. An *Operation LSN* is unchanged from before. But an *abLSN* is more complicated, and hence the resulting <= test is more complicated as well.

**Abstract LSNs:** We need to capture precisely which operations have results captured in the state of a page. We define our *abLSN* as accurately capturing every operation that has been executed and included in the state of the page. More precisely, it needs to indicate which operations' results are not included on the page. Our *abLSN* consists of a low water *LSNlw*, whose value is such that no operation with an *LSN < =LSNlw* needs to be re-executed. We augment *LSNlw* with the set *{LSNin}* of LSNs of operations greater than *LSNlw* whose effects are also included on the page. Thus we have *abLSN = <LSNlw, {LSNin}>*. An operation with *LSNi* has results captured in the page with *abLSN* when *LSNi <= abLSN* where <= is defined as:

*LSNi <= abLSN iff LSNi<=LSNlw or LSNi in {LSNin}*

**Establishing LSNlw**: How can the DC know that a particular value is suitable as *LSNlw*? This means that the DC would have already performed every operation with lower LSN which might be applicable on that page. If DC has a pending unapplied operation with a lower LSN, it knows this, but because of multithreading, operations can come to the DC out of LSN order. Thus the DC can't determine by itself which operations are not yet applied. However, the TC knows which LSNs were generated for operations, and which have definitely been performed. So, from time to time, the TC will send the DC *LWM* such that the TC has received replies from the DC for all operations with LSNs up to *LWM*.

The DC can use the TC supplied *LWM* in any of its cached pages as the *LSNlw* for the page. Simultaneously, the DC can discard from the *abLSN* for the page any element of *{LSNin}* such that *LSNin <= LSNlw*.

**Page Sync:** During normal execution, we do not need to keep *abLSN* in the page itself, as long as it is available in volatile memory outside the page, to be tested as required. However, when the page is flushed to disk, the *abLSN* must be made stable atomically with the page. Traditionally, this is done by including LSN information in the page itself, and we focus on this approach here. We call this step a page sync, and require that all pages be synced before being written to volatile storage.

There are two distinct ways that pages can be synced, and some combination of the two is also possible. When a page is to be flushed, we could follow any of these algorithms:

1. We refuse to execute operations on the page with LSN's greater than the highest valued *LSNin*. Eventually, the *LSNlw* sent by the TC will equal or exceed every *LSNin*, at which time we can set *abLSN* for the page to *LSNlw*. This delays the page flush.

2. We include the entire existing *abLSN* on the page. This takes up more storage on the page than a single LSN would.

3. We wait until the number of LSNs in *{LSNin}* is reduced to a manageable size using a TC supplied *LSNlw*, and then include the *abLSN* on the page which is then flushed.

## 5.2 System Transaction Execution Order

### 5.2.1 Current Technique

Most modern database systems exploit some form of atomic action to provide recovery for access method structure modifications [16, 20]. Indeed Microsoft SQL Server uses a variant of multi-level recovery in which system transactions encapsulate structure modifications. The characteristic of existing system transactions [15] is that like open nested transactions, system transactions are redone in precisely their original execution order. Undo recovery is done in two steps. First, incomplete system transactions are undone, then user transaction level transactions are undone. This is the usual multi-step undo done for multi-level transactions and it ensures that logical user transaction undo operations find a search structure that is well formed when they are executed.

### 5.2.2 Our New Technique

When we split the kernel, it is the DC that handles all page related operations, including all structure modifications to an index structure. These structure modification operations need to be atomic and recoverable. Microsoft SQL Server uses latching and system transactions for this. Because this is now done by the DC, both latching and the logging needed for system transactions must be done by the DC. Further, the DC will use its own LSNs *(dLSNs)* to make structure modification recovery idempotent. That is, each page should contain both *dLSN* (indicating which structure modifications are reflected in this page) and *abLSN* as described in 5.1.2.

Splitting the kernel requires that the TC submit logical redo as well as logical undo operations to the DC. Hence, indexes maintained by the DC need to be well-formed before considering any logical redo sent by TC. That is, the DC needs to make its search structures well-formed by completing any redo and undo of system transactions from the DC-log, prior to the TC executing its redo recovery. This moves system transaction recovery ahead of all TC level recovery. This change in the order of recovery means we need to manage LSN information correctly in order to indicate what operations (both from the DC-log and from the TC-log) are reflected in the page. To make this concrete we discuss the system transactions involved in page splits and page deletions in a B-tree.

**Page Splits:** Page splits make additional storage available to a B-tree. The DC-log has (among other log records) an entry that records the creation of the new page, and an entry that records the removal of keys from the pre-split page. When these DC-log events are moved forward during recovery, the page split is executed earlier in the update sequence relative to the TC operations that triggered the split. Repeat-history recovery can be made to work for this case.

1. The DC-log record for the new page needs to capture the page's *abLSN* at the time of the split since the log record for the new page contains the actual contents of the page.

2. The DC-log record for the pre-split page need only capture the split key value. Whatever version of that page exists on stable storage, its *abLSN* captures the state of this page. And we can use that *abLSN* validly for this page, whether we find it in a state prior to or later than the split.

**Page Deletes/Consolidates:** When a page of an index structure is deleted, the search range for the page is logically consolidated with an adjacent page of the index structure. Such page deletes are moved forward in their execution during recovery as the DC-log is recovered before TC recovery, but this seriously complicates recovery. Page deletes reduce the amount of space available to the index structure. A consolidation that happens early may find that the versions of the pages involved during recovery don't fit into a single page. When the DC executes internal *system transactions that do not commute* with previously executed TC-generated logical operations, the DC must provide a recoverable state that includes these prior operations (e.g., by generating a physical log record that encompasses the non-commutative prior operations). Thus, we can make an entry in DC-log for the deletion of the page whose space is to be returned to free space; this log record can be logical, indicating that the page is no longer needed. But when we produce a DC-log entry for the consolidated page which inherits the deleted page's key range and perhaps the remaining records in that range, we use a physical DC-log record that captures the entire page including using an abLSN for the consolidated page that is the maximum of *abLSNs* of the two pages; redoing the consolidation amounts to giving the consolidated page the contents and key-range that it had immediately after the consolidation originally happened. That is, this logging/recovery technique forces the delete to keep its position in the execution order wrt TC-submitted operations which are on the key range of the consolidated page. Such physical logging of a consolidated page is more costly in log space than the traditional logical system transaction for page deletes. But page deletes are rare, so the extra cost should not be significant.

## 5.3 Partial Failures

### 5.3.1 Current Technique
There are no current database techniques for this, as this situation cannot arise. Failures in a monolithic database kernel are never partial. Log and cache manager fail together.

### 5.3.2 Our New Technique
By splitting a database kernel, we need to face the possibility that TC and DC fail independently. The complete failure of both TC and DC returns us to the current fail-together situation and requires no new techniques. Now, consider separate (and hence partial) failures.

**DC Failure:** When the DC fails, it loses its volatile (in-cache) state. The database state in the DC reverts to the state captured on stable storage. Once the TC has been notified, it resends operations forward from the redo scan start point (as indicated in the checkpoint). The DC re-applies any of these operations which are missing from the stable state. This is conventional recovery.

An important point in an unbundled design is that the TC chooses the redo scan start point based on which operations have their idempotence-guarantee terminated, because the DC has checkpointed all these operations' effects; communicating from the DC to the TC that this has happened requires an extension to the interface between the components.

**TC Failure:** When the TC fails and loses its log buffers while the DC continues to run normally, the TC needs a way of *resetting the state of the DC* to an appropriate earlier state. The problem is that the TC loses the tail of its log that had not been forced to stable storage, and some of these operations may have been already performed in a DC. Note that such pages can only be in a DC's cache; the causality principle enforces that no such pages are stable in a DC. That is, the DC cache may contain pages which reflect the effects of TC operations that have been lost. This must be reversed *before* the TC resends operations from its stable log to be re-applied in a DC.

We can proceed in a number of ways to reset the DC state to an earlier appropriate state. One way is to turn a partial failure into a complete failure. This drops all pages from the DC cache and permits conventional recovery to work. However, there is no need to be this draconian. A more efficient method is to drop from the cache only those pages that contain the results of operations that have been lost. Once we do this, the TC can begin resending operations; the DC re-applies each, perhaps fetching the relevant page(s) from disk if they are no longer in the DC's cache. The pages that the DC must drop from its cache to reset state correctly are exactly the pages whose *abLSNs* include operations that are later than $LSN_{st}$, the largest LSN on the TC stable log.

## 6. MULTIPLE TC'S FOR A DC
It is possible to permit more than one TC to update data at a given DC. So long as the records of each application are disjoint (data is logically partitioned), having multiple TCs accessing data at a given DC can be supported, as the invariant that no conflicting operations are active simultaneously can be enforced separately by each TC. This does impose additional requirements on such a DC, however.

## 6.1 DC Requirements

### 6.1.1 Multiple Abstract LSNs
A DC supporting multiple TCs must be prepared to provide idempotence for each of the TCs. Since TCs do not coordinate how they organize and manage their logs, the *LSN*s from each TC need to be tracked separately by the DC. Thus, each page would needs to include an *abLSN* for each TC that has data on the page. However, pages with data from only a single TC continue to have only one *abLSN*. So, only on pages containing data from multiple TCs would extra *abLSN*'s be needed.

### 6.1.2 Resetting the Database Buffer
When a TC crashes, it may lose the log records for requests that it sent to a DC. The DC must be able to reset the pages that it has in its volatile cache (the changes cannot have propagated to the disk). We have already described this for single TC sending requests to a DC. It is highly desirable that a DC be able to reset pages that are affected by a TC crash so that only the failing TC need resend requests and participate in recovery.

The DC needs to reset pages where the *abLSN* of the failed TC has captured operations that were not on the stable log when the failed TC crashed. Identifying these pages is easy since the pages all have *abLSN* s for every TC with data on the page. However, unlike before, we cannot simply replace such a page with the disk version of the page and then ask the failed TC to resend the appropriate requests. The disk version of the page may also not contain changes produced by non-failing TCs. Such a replacement from disk would require that the other TCs with updates that are removed replay their logs to restore these pages. This is exactly what we want to avoid.

We need to identify the data on each page that is associated with our failed TC. We continue to not want to associate an *LSN* with each record, though that is less of a hardship with multiple TCs. However, we expect most pages to have updates from a single TC, so we want to optimize for this case. To reset the pages containing lost updates of our failed TC, we need to associate the failed TC's *abLSN* on the page with the data to which it applies. One way to accomplish this is to link the records related to a TC to the single occurrence of the TC's *abLSN* on the page. Such links could, e.g., be two byte offsets that chain the records together.

A page reset then consists of replacing the records on the page updated by a failed TC with the records from the disk version of the page. Records updated by other TCs would not be reset.

## 6.2 Sharing Data Among TCs

Recall that operations executing concurrently at a DC must not conflict. Hence, if we can limit the types of requests that multiple TCs execute at a given DC to ones that are non-conflicting, we can permit shared access to the data managed by a DC. In this case, the assignments of logical portions of the data to different TCs need not be disjoint. We cannot permit arbitrary sharing, but some types of sharing can be provided, so long as the reads are at low isolation levels. We first describe types of TC shared access to data that can be supported without any additional mechanism. We then show how a good bit more sharing can be supported via versioned data.

### 6.2.1 Non-versioned Data

**Read-Only:** All reads commute, regardless of their source. So it is possible for multiple TCs to share read-only data with each other without difficulty. The data read will be transaction consistent because no TC can change the data.

**Dirty Reads:** It is sometimes possible to share read and write access to mutable data. Dirty reads, where uncommitted data may be read, do not require any locking for reads. A writer may access and update data ("make it dirty") at any time without conflicting with a dirty read. Because a DC provides operation atomicity, a reader of dirty data will always see "well formed" data, though this data may be from uncommitted transactions. Dirty data can disappear should the updating transaction abort. Further, it can be modified subsequently, before its transaction commits. However reading dirty data can sometimes be useful despite these caveats.

Note that the above functionality requires no special DC knowledge or implementation.

### 6.2.2 Versioned Data

**Read Committed Access:** With versioned data, we can permit TCs that update disjoint data partitions at a DC to perform "read committed" reads of data updated by other TCs. With versioned data, an update produces a new uncommitted version of the record, while continuing to maintain an earlier before version. To provide an earlier version for inserts, one can insert two versions, a before "null" version followed by the intended insert.

When an updating TC commits the transaction, it sends updates to the DC to eliminate the before versions, making the later versions the committed versions. Should the transaction abort, the TC sends operations to the DC instructing it to remove the latest versions that were updated by the transaction.

A reader from another TC that encounters a record with a before version reads the before version. If it encounters a record without a before version, it reads this single version. A TC executing a transaction can be permitted to see its own updates on its own disjoint updatable partition while also reading committed data from other TCs. To do this requires that it issue a different flavor of read for its own partition of data.

An important characteristic of this approach is that there is no classic (blocking) two phase commit protocol in this picture. Once the TC decides to commit, the transaction is committed everywhere and it is guaranteed that the earlier before versions of its updates will eventually be removed. An updating TC is only blocked when it is actually down, in which case, none of its data is updatable in any event. The situation is similar when an updating TC decides to abort. Readers are never blocked. Interestingly, this is non-blocking exactly because "read committed" access is being used with versioning.

## 6.3 Cloud Sharing Scenario

An example that captures some of the kinds of sharing of data across TCs that is desired in a cloud setting is an online movie site that tracks information about movies and allows users to write reviews. The fundamental problem here is that we want to cluster every review with both its reviewer and with the movie it discusses. That permits high-performance clustered access for reading the reviews of a given movie (the most common operation in the system), as well as high-performance clustered access to a user and all her reviews. Unclustered access in the cloud is enormously more expensive, requiring access to a potentially very large collection of computers. However, at such a site the most common update transactions involve a single user's data (reviews, profile, favorites, etc). As such it is desirable to avoid distributed transactions when users update their data and add reviews while still providing full transaction semantics across updates that span machines in the cloud.

There are four common transaction workloads to consider:

1. W1: obtain all reviews for a particular movie
2. W2: add a movie review written by a user
3. W3: update profile information for a user
4. W4: obtain all reviews written by a particular user

There are four tables to support these workloads:

1. Movies (primary key MId): contains general information about each movie. Supports W1.
2. Reviews (primary key MId, UId) contains movie reviews written by users. Updated by W2 to support W1.
3. Users (primary key UId): contains profile information about users. Updated by W3.
4. MyReviews (primary key UId, MId): contains a copy of reviews written by a particular user. Updated by W2 to support W4. Effectively this table is an index in the physical schema since it contains redundant data from the Reviews table.

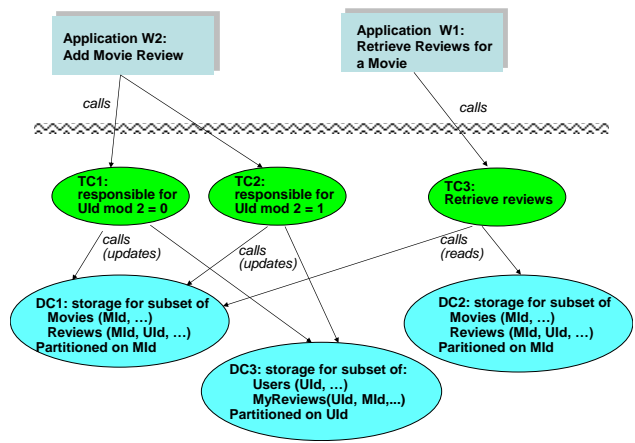

**Figure 2: TC and DC Partitioning**

Figure 2 illustrates how data and transactions can be partitioned across TCs and DCs to achieve the goals of running the above workload without distributed transactions and without a query needing to access more than two machines to retrieve the desired data. Users and their workload (W2-W4) are partitioned among TCs, in this case TC1 and TC2. These TCs have full access rights to all information about a user in the Users table and also have access rights to insert ("post") reviews *by that user* in the Reviews table. No one else has the right to post movie reviews by a particular user at any movie, so this is also a disjoint partitioning. Clearly, the updating TC can also read the user information as it has full access rights to it. The Users table and MyReviews table may also be partitioned by user across DCs and this illustration shows DC3 containing a such a partition.

With this partitioning, TC1 can add a movie review for a user by updating DC1 to insert the review in the Reviews table and DC3 to insert it in the MyReviews table. The transaction is completely local to TC1. Users can also obtain all of their reviews (W4) by simply querying a single partition of the MyReviews table.

We also wish to enable TC3 to read all of the reviews for a movie in a single query (W1). Given that a movie may have a large number of reviews and that requests to read the reviews will be much more common than adding reviews, it is critical to cluster reviews with their corresponding movies on a single DC. To achieve this clustering the Movies and Reviews tables are partitioned by movie onto DC1 and DC2.

In this example, TC3 requires shared access. We cannot use "read only" access since we are permitting the data involved to be updated. We can solve this problem without versioning if dirty reads are acceptable, as they do not conflict with access by updaters. With versioning, we can provide *read committed* access as well, since such versioned reads do not conflict with updates. We also see potential for providing snapshot isolation [4] and perhaps selectively strengthening it into serializability as needed by the applications [5].

Thus, with shared (non-conflicting) access, we can support some important scenarios that, on the surface look impossible to provide.

## 7. Conclusion

This paper suggests a paradigm shift in the way transactional recovery and concurrency control are provided in data management platforms. We have worked out our proposal in sufficient detail to be convinced that it is implementable and reasonably efficient. However, compared to a traditional storage kernel with integrated transaction management, our unbundling approach inevitably has longer code paths. Our working hypothesis is that this is justified by the flexibility of deploying adequately-grained cloud services. In addition, we speculate about possible throughput gains on multi-core processors: with more compact code for separate TC and DC, and the ability to instantiate each multiple times with configurable numbers of threads, we hope for more effective use of cores and better cache hit rates. Ongoing work is needed to demonstrate these effects.